\documentclass[preprint,showpacs,aps]{revtex4}
\usepackage{graphicx}

\begin{document}

\title{Half-lives of $\alpha$-emitters approaching the N=Z line}
\author{Chang Xu$^{1}$ and Zhongzhou Ren$^{1,2,3} \email{ zren@nju.edu.cn}$ }
%\email{ zren@nju.edu.cn}
\address{$^{1}$Department of Physics, Nanjing University,
Nanjing 210008, China\\
$^{2}$Center of Theoretical Nuclear Physics, National Laboratory
of Heavy-Ion Accelerator, Lanzhou 730000, China \\
$^{3}$CPNPC,  Nanjing University, Nanjing 210008, China}

\begin{abstract}

The half-lives of newly observed $\alpha$-emitters $^{105}$Te and
$^{109}$Xe [Seweryniak \textit{et al.}, Phys. Rev. C \textbf{73},
061301(R) (2006); Liddick \textit{et al.}, Phys. Rev. Lett.
\textbf{97}, 082501 (2006)] are investigated by the
density-dependent cluster model. The half-lives of
$\alpha$-emitters close to the N=Z line are also studied in a
unified framework where the influence of the nuclear deformation
is properly taken into account. Good agreement between model and
data is obtained and some possible $\alpha$-emitters are suggested
for future experiments.

\end{abstract}


\maketitle

One of the important subjects in nuclear physics is the
exploration of the stability of nuclei against spontaneous
radioactivity. Experiments have shown that a large number of
nuclei are unstable to $\alpha$ emission, which is a primary decay
mode for medium and heavy nuclei in the mass table \cite{aud}.
Since the discovery of $\alpha$ radioactivity, the mass region
dominated by $\alpha$-decay has been continuously broadened owing
to the rapid development of experimental techniques.
Theoretically, a very interesting question concerned arises: Which
are the lightest and heaviest $\alpha$-emitters? In fact, the
available heaviest $\alpha$-emitters synthesized by cold and hot
fusion reaction methods have been studied extensively
\cite{hof,oga}. The theoretical results of many different
$\alpha$-decay models were in good agreement with the experimental
data \cite{buc,roy,del,del2,den,gam,moh,dua,cho,xu1}. Useful
predictions on even heavier $\alpha$-emitters were also given in a
number of theoretical works. Significant progress has been made
toward the synthesis of the heaviest $\alpha$-emitter from both
the experimental and theoretical sides. In contrast,
$\alpha$-emitters at the opposite end of the mass table have
received little attention \cite{aud}. Usually the light mass
$\alpha$-emitters are very close to the N=Z line, which can
provide unique information of nuclear structure as compared with
the neutron-rich ones. Thus it is also important to investigate
the decay properties of the light mass $\alpha$-emitters and to
make reliable theoretical predictions. Reaching limit of the
lightest $\alpha$-emitter is, at the same time, a very interesting
objective for future experiments. Recently, the $\alpha$-decay of
$^{105}$Te with a decay energy of Q$_{\alpha}$ = 4900(50) keV and
a half-life of T$_{1/2}$ = 0.70($-$0.17 + 0.25) $\mu$$s$ was
reported by Seweryniak \textit{et al.} \cite{sew,lid}. This newly
found $\alpha$-emitter is the lightest one observed experimentally
so far. It belongs to a special class of $\alpha$-emitters
occurring in the mass region A=105$-$114 with N/Z ratio
$\approx$1.0 \cite{aud}.

In this Brief Report, we systematically investigate the
$\alpha$-decay of $^{105}$Te and other light mass
$\alpha$-emitters ($^{106}$Te$-$$^{114}$Ba) in the framework of
the density-dependent cluster model (DDCM). The density-dependent
cluster model is based on the microscopic double-folding potential
with the M3Y nucleon-nucleon interaction, which correctly includes
the low-density behavior of the nucleon-nucleon interaction and
guarantees the antisymmetrization of identical particles in the
$\alpha$-cluster and in the core \cite{sat,ber}. In DDCM the
ground-state of the parent nucleus is assumed to be an
$\alpha$-particle interacting with a deformed core that has an
axially symmetric deformation. The influence of the core
deformation on the double-folding potential is properly taken into
account by the multipole expansion method. The depth of the
nuclear potential is determined separately for each decay to
generate a quasibound state. The penetration probability of the
$\alpha$-particle through the deformed Coulomb barrier is obtained
by a careful averaging procedure along different orientation
angles \cite{ste}. The only free parameter in DDCM is the
preformation factor which is chosen to fit the experimental
half-lives. The value of the preformation factor in DDCM is
consistent with both the experimental facts and the microscopic
calculations. The details of DDCM can be found in Ref.\cite{xu2}.

The $\alpha$-decay of the isotopic chains of Te, I, Xe, Cs, and Ba
forms a special light mass region around $^{100}$Sn and we call it
as `light island' of $\alpha$-decay for convenience. The maximum
mass number of the observed $\alpha$-emitters in the `light
island' is A=114 ($^{114}_{\,\,56}$Ba) \cite{maz}, whereas the
minimum mass number of the $\alpha$-emitters outside the `light
island' is A=144 ($^{144}_{\,\,60}$Nd) \cite{aud}. We know that
the $\alpha$-emitters heavier than $^{114}_{\,\,56}$Ba are all
very neutron-rich. This differs from the $\alpha$-emitters in the
`light island' which approximately consist of equal numbers of
protons and neutrons.
\begin{figure}[htb]
\centering
\includegraphics[width=9cm]{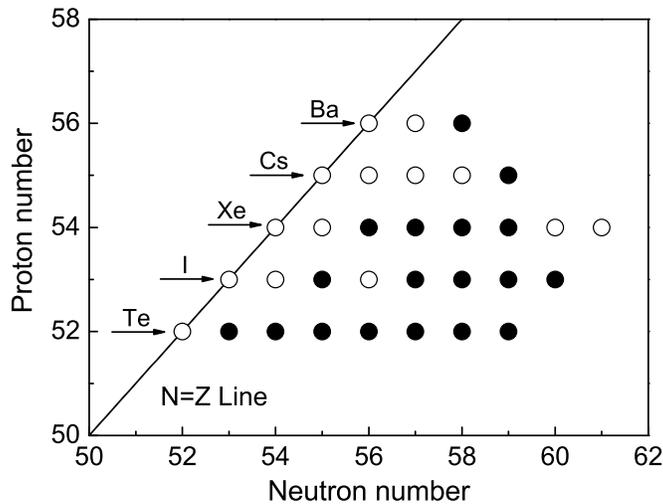}
\caption{Alpha emitters close to the N=Z line. }
\end{figure}
In Fig.1, we plot the `light island' of Te, I, Xe, Cs, and Ba
isotopes. The X and Y axes are the neutron and proton numbers,
respectively. The black circles denote the $\alpha$-emitters with
experimental data and the white circles denote the possible
$\alpha$-emitters which have not been observed in experiment yet.
It is seen from Fig.1 that the presently found $\alpha$-emitter
$^{105}$Te is the one closest to the N=Z line. It consists of a
combination of 52 protons and 53 neutrons. Experimentally it is
also possible to observe the first $\alpha$-emitter with equal
proton and neutron numbers.

We perform a systematic calculation on the $\alpha$-emitters in
the `light island' by the deformed version of the
density-dependent cluster model \cite{xu1}. The detailed numerical
results are given in Table I. In Table I, the parent and daughter
nuclei are listed in the first column. The $\alpha$-decay energy
is listed in the second column. The deformation parameters in
column 3 and column 4 are taken from M\"{o}ller \textit{et al.}
\cite{mo1}. The experimental $\alpha$-decay branching ratio is
given in column 5 \cite{web}. The experimental partial half-lives
and two sets of calculated results are listed in the last three
columns, respectively.

Besides the available $\alpha$-emitters, we also give in Table I
the suitable candidates of $\alpha$-emitters in the light mass
region for future experiments (see also Fig.1). To make reliable
predictions on the half-lives of these possible $\alpha$-emitters,
the $\alpha$-decay energies used in calculations should be chosen
very carefully. Through a detailed analysis on the experimental
data, it is found that there exists a linear relationship for the
experimental $\alpha$-decay energies of a certain kind of Te, I
and Xe isotopes. For instance, Fig.2 plots the experimental
$\alpha$-decay energies of $^{106}$Te, $^{108}$Te and $^{110}$Te
against the mass numbers (A).
\begin{figure}[htb]
\centering
\includegraphics[width=7.5cm]{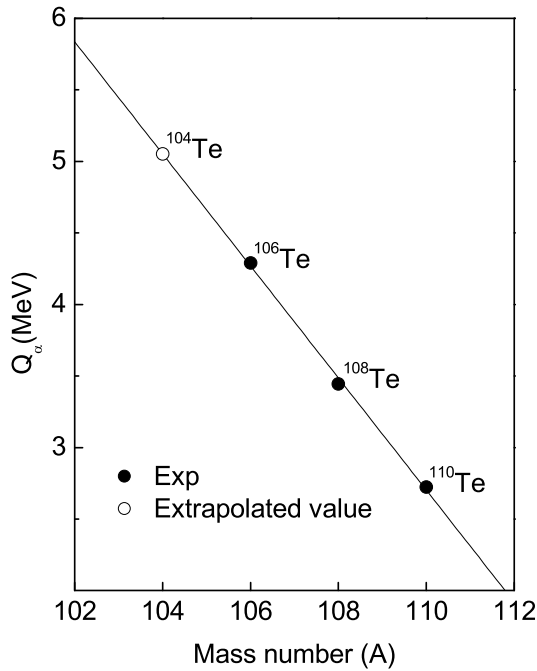}
\caption{Extrapolation of $\alpha$-decay energy for $^{104}$Te.}
\end{figure}
It is clearly seen from Fig.2 that the three experimental points
(black circles) of $^{106}$Te, $^{108}$Te and $^{110}$Te are in a
perfect line. It is expected that the $\alpha$-decay energy of
$^{104}$Te (white circle) also lies on this line and the
extrapolated $\alpha$-decay energy for $^{104}$Te, as well as for
other predicted $\alpha$-emitters should have good reliability.
For the decay chains of Cs and Ba, the theoretical $\alpha$-decay
energies from the finite-range droplet model (FRDM) are used in
calculations because only a single $\alpha$-emitter in each chain
is observed in experiment \cite{mo2}. We note that the predicted
$\alpha$-decay energies of Cs and Ba isotopes from FRDM are very
accurate as compared to the experimental ones \cite{mo2}. For
example, the theoretical decay energy of $^{114}$Ba from FRDM is
3.550 MeV, which is very close to the experimental value
(Q$_{\alpha}$=3.540 MeV) \cite{maz}.

The quadrupole and hexadecapole deformations are generally not
large for the $\alpha$-decay chains of Te, I, Xe, Cs, and Ba (see
Table I). We properly take into account the influence of nuclear
deformation on the partial half-lives in our calculations. From
Table I, it is seen that the magnitude of the nuclear deformation
increases with the increasing neutron number for each isotopic
chain. The largest deformation occurs for the $\alpha$-emitter
$^{114}$Ba, which has a quadrupole deformation of $\beta_2$=0.169
and a hexadecapole deformation of $\beta_4$=0.052. Similarly, the
experimental $\alpha$-decay branching ratios in Table I also
reveal a close dependence on the neutron number (N). Usually the
$\alpha$ emission is a dominating decay mode on the proton-rich
side of each isotopic chain, \textit{i.e.}
$b_{\alpha}$\%$\approx$100\%. On the neutron-rich side, the
$\alpha$-decay mode of the nuclei competes with the $\beta$-decay
process and the value of the $\alpha$-decay branching ratio is
much smaller ($b_{\alpha}$\%$\ll$100\%). Moreover, the proton
radioactivity also can become a very important decay mode for
several nuclei, such as $^{109}$I, $^{112}$Cs and $^{113}$Cs, but
the observation of $\alpha$ emission of the three nuclei is still
feasible.

In Table I, we can see that the experimental partial
$\alpha$-decay half-lives range from 7.0$\times10^{-7}$ to
2.0$\times10^{7}$ seconds. This is a very large variation for the
$\alpha$-decay lifetime. However, the first set of calculated
partial half-lives by the deformed version of DDCM is in good
agreement with the available experimental data. For example, the
theoretical half-life of $^{105}$Te is
$\textrm{T}_{\alpha}{(\textrm{Cal1})}$=0.54 $\mu$s, which agrees
well with the experimental one (T$_{\alpha}$(Exp)=0.70 $\mu$s)
\cite{sew}. The experimental half-lives of many $\alpha$-emitters
in Table I are reproduced within a factor of 3 by DDCM.
Experimental and theoretical values for $^{114}$Ba deviate by a
factor of 5, which is the largest among all $\alpha$-emitters. One
possible reason of this abnormal deviation is that the theoretical
deformation parameters from FRDM are relatively too small for
$^{114}$Ba \cite{mo1}. When larger values of the quadrupole and
hexadecapole deformations are used in calculations, the agreement
between model and data can be greatly improved for $^{114}$Ba
\cite{maz}.
\begin{figure}[htb]
\centering
\includegraphics[width=10.5cm]{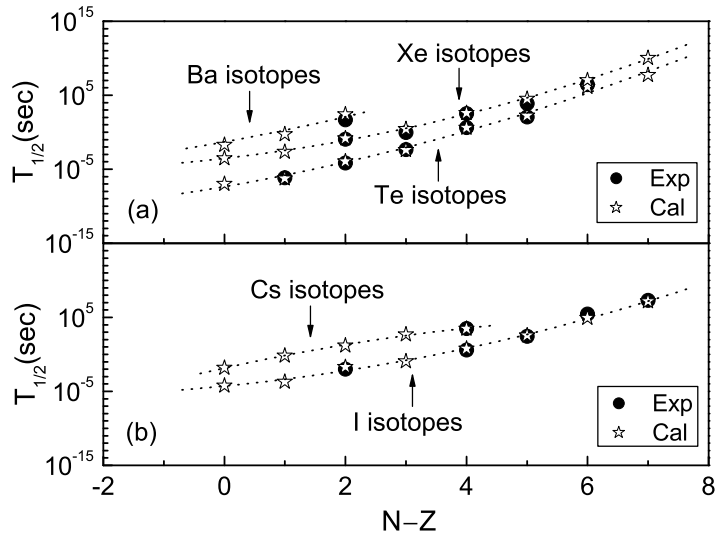}
\caption{Experimental and theoretical $\alpha$-decay half-lives
for isotopes of Te, I, Xe, Cs, and Ba.}
\end{figure}
In Fig.3, we give a direct comparison of the experimental and
theoretical $\alpha$-decay half-lives of these nuclei
($\textrm{T}_{\alpha}{(\textrm{Exp})}$ and
$\textrm{T}_{\alpha}{(\textrm{Cal1})}$). The even-Z isotopes of
Te, Xe, and Ba are plotted in the upper part, and the odd-Z
isotopes of I and Cs in the lower part of Fig.3. It is seen from
Fig.3 that the experimental points follow the theoretical curve
very well. This shows that DDCM has good accuracy in the light
mass region. The predicted $\alpha$-emitters are also given in
Fig.3. We consider that these predicted $\alpha$-emitters are
within the range of current experimental capabilities. Other
predicted $\alpha$-emitters have either an extremely short
lifetime or a very weak $\alpha$-decay branching ratio, which are
not included in our calculations. Because of the success of DDCM
for the available experimental data, the present exploration to
unknown mass region is necessary and useful for experiments.

Although the first set of calculated results
($\textrm{T}_{\alpha}{(\textrm{Cal1})}$) is very close to the
experimental data, the overall agreement can be further improved
by taking into account the degree of $\alpha$-clusterization in
DDCM. Delion and co-workers systematically analyzed the
$\alpha$-clustering effect in heavy and superheavy nuclei
\cite{del,del2} and they pointed out the suppression of
$\alpha$-clusterization process with increasing proton-neutron
asymmetry along the isotopic chains \cite{del2}. To improve the
agreement between experiment and theory, we therefore use the
isospin-dependent preformation factor
P$_{\alpha}$=c$_1$+c$_2$(N$-$Z) instead of the constant one
\cite{xu2} for each kind of nuclei [e.g., a linear dependence
P$_{\alpha}^{ee}$=0.73$-$0.09$\times$(N$-$Z) for the even-even
nuclei]. As expected, the corresponding theoretical lifetimes
($\textrm{T}_{\alpha}{(\textrm{Cal2})}$) show a significantly
better agreement with the experimental data. The root-mean-square
deviation reduces from 0.319 to 0.242 for the available
$\alpha$-emitters. Thus the experimental data are reproduced very
well by including the nuclear structure effect of
$\alpha$-clusterization in DDCM.

In summary, we carried out a systematic study of the
$\alpha$-decay half-lives of Te, I, Xe, Cs, and Ba isotopes by the
deformed version of the density-dependent cluster model. This is a
test on the reliability of DDCM for the very light mass region
just above $^{100}$Sn. For the newly observed $\alpha$-emitters
$^{105}$Te and $^{109}$Xe, the calculated partial half-lives are
in good agreement with the experimental data. For other light mass
$\alpha$-emitters ($^{106}$Te$-$$^{114}$Ba), the experimental
half-lives are also reproduced very well. In addition, we also
give the theoretical half-lives of some possible $\alpha$-emitters
for future experiments which includes five predicted
$\alpha$-emitters consisting of equal numbers of protons and
neutrons. It is very interesting to compare the present
theoretical predictions with the experimental observations in
future.

\

Notes added: After the submission of this article, we noticed that
one of our predicted $\alpha$-emitters $^{109}$Xe was reported by
Liddick \textit{et al.} \cite{lid}, and we added the new results
of the favored $\alpha$-transition ($\frac{7}{2}^+ \rightarrow
\frac{7}{2}^+$) of $^{109}$Xe in the revised version (see Table
I). For the $\alpha$-transition to the ground-state of $^{105}$Te
($\frac{7}{2}^+ \rightarrow \frac{5}{2}^+$), DDCM also yields a
theoretical partial lifetime (Q$_{\alpha}(\textrm{Exp})$=4.217MeV,
$\textrm{T}_{\alpha}{(\textrm{Cal2})}$=17ms) in good agreement
with the experiment one
($\textrm{T}_{\alpha}{(\textrm{Exp})}$=21ms).

\

This work is supported by National Natural Science Foundation of
China (No.10125521, No.10535010) and by 973 National Major State
Basic Research and Development of China (No.G2000077400).

\begin{table*}[htb]
\centering \caption{The comparison of experimental and theoretical
$\alpha$-decay partial half-lives of Te, I, Xe, Cs, and Ba
isotopes (in seconds).}
\begin{tabular}{clccclll}
\hline
\hline
         $\textrm{Nuclei}$
        &\hspace{0.1cm} $\textrm{Q}_{\alpha}{(\textrm{MeV})}$
        &\hspace{0.35cm} $\beta_2$ \hspace{0.35cm}
        &\hspace{0.35cm} $\beta_4$ \hspace{0.35cm}
        &$b_{\alpha}\%{(\textrm{Exp})}$
        &$\textrm{T}_{\alpha}{(\textrm{Exp})}$
        &$\textrm{T}_{\alpha}{(\textrm{Cal1})}$\hspace{0.5cm}
        &$\textrm{T}_{\alpha}{(\textrm{Cal2})}$\\
\hline
$^{104}$Te$\rightarrow$$^{100}$Sn+$\alpha$   &\hspace{0.4cm} 5.053$^{*}$  & 0.009 & -0.015 &            &                    & 9.8$\times10^{-8}$  & 5.1$\times10^{-8}$\\[-0.15cm]
$^{105}$Te$\rightarrow$$^{101}$Sn+$\alpha$   &\hspace{0.4cm} 4.900        & 0.027 &  0.024 & 100\%      & 7.0$\times10^{-7}$ & 5.4$\times10^{-7}$  & 7.6$\times10^{-7}$\\[-0.15cm]
$^{106}$Te$\rightarrow$$^{102}$Sn+$\alpha$   &\hspace{0.4cm} 4.290        & 0.009 & -0.015 & 100\%      & 7.0$\times10^{-5}$ & 1.1$\times10^{-4}$  & 7.6$\times10^{-5}$\\[-0.15cm]
$^{107}$Te$\rightarrow$$^{103}$Sn+$\alpha$   &\hspace{0.4cm} 4.008        & 0.018 & -0.015 &  70\%      & 4.4$\times10^{-3}$ & 3.8$\times10^{-3}$  & 3.6$\times10^{-3}$\\[-0.15cm]
$^{108}$Te$\rightarrow$$^{104}$Sn+$\alpha$   &\hspace{0.4cm} 3.445        & 0.018 &  0.016 &  49\%      & 4.3$\times10^{0}$  & 3.9$\times10^{0}$   & 4.0$\times10^{0}$\\[-0.15cm]
$^{109}$Te$\rightarrow$$^{105}$Sn+$\alpha$   &\hspace{0.4cm} 3.230        & 0.026 &  0.009 &   4\%      & 1.2$\times10^{2}$  & 1.7$\times10^{2}$   & 1.2$\times10^{2}$\\[-0.15cm]
$^{110}$Te$\rightarrow$$^{106}$Sn+$\alpha$   &\hspace{0.4cm} 2.723        & 0.027 &  0.016 &  0.00067\% & 2.8$\times10^{6}$  & 1.3$\times10^{6}$   & 2.6$\times10^{6}$\\[-0.15cm]
$^{111}$Te$\rightarrow$$^{107}$Sn+$\alpha$   &\hspace{0.4cm} 2.576        & 0.045 &  0.001 & $\ll$100\% &                    & 5.2$\times10^{7}$   & 3.0$\times10^{7}$\\[-0.10cm]
\hline
$^{106}$I$\rightarrow$$^{102}$Sb+$\alpha$    &\hspace{0.4cm} 4.579$^{*}$  & 0.054 &  0.034 &            &                    & 6.1$\times10^{-5}$  & 3.1$\times10^{-5}$\\[-0.10cm]
$^{107}$I$\rightarrow$$^{103}$Sb+$\alpha$    &\hspace{0.4cm} 4.400$^{*}$  & 0.054 &  0.034 &            &                    & 1.9$\times10^{-4}$  & 2.7$\times10^{-4}$\\[-0.15cm]
$^{108}$I$\rightarrow$$^{104}$Sb+$\alpha$    &\hspace{0.4cm} 4.034        & 0.081 &  0.051 & 100\%      & 1.0$\times10^{-2}$ & 1.9$\times10^{-2}$  & 1.2$\times10^{-2}$\\[-0.15cm]
$^{109}$I$\rightarrow$$^{105}$Sb+$\alpha$    &\hspace{0.4cm} 3.835$^{*}$  & 0.081 &  0.051 &            &                    & 1.1$\times10^{-1}$  & 1.1$\times10^{-1}$\\[-0.15cm]
$^{110}$I$\rightarrow$$^{106}$Sb+$\alpha$    &\hspace{0.4cm} 3.580        & 0.099 &  0.052 & 17\%       & 3.8$\times10^{0}$  & 6.1$\times10^{0}$   & 5.7$\times10^{0}$\\[-0.15cm]
$^{111}$I$\rightarrow$$^{107}$Sb+$\alpha$    &\hspace{0.4cm} 3.270        & 0.098 &  0.052 & 0.09\%     & 2.8$\times10^{2}$  & 3.6$\times10^{2}$   & 2.6$\times10^{2}$\\[-0.15cm]
$^{112}$I$\rightarrow$$^{108}$Sb+$\alpha$    &\hspace{0.4cm} 2.990        & 0.107 &  0.044 & 0.0012\%   & 2.9$\times10^{5}$  & 8.7$\times10^{4}$   & 1.4$\times10^{5}$\\[-0.15cm]
$^{113}$I$\rightarrow$$^{109}$Sb+$\alpha$    &\hspace{0.4cm} 2.705        & 0.107 &  0.044 &3.31$\times10^{-5}$\%& 2.0$\times10^{7}$  & 1.5$\times10^{7}$   & 0.9$\times10^{7}$\\[-0.10cm]
\hline
$^{108}$Xe$\rightarrow$$^{104}$Te+$\alpha$   &\hspace{0.4cm} 4.440$^{*}$  & 0.045 &  0.016 &            &                    & 2.9$\times10^{-4}$  & 1.5$\times10^{-4}$\\[-0.10cm]
$^{109}$Xe$\rightarrow$$^{105}$Te+$\alpha$   &\hspace{0.4cm} 4.067        & 0.053 &  0.017 &            & 5.0$\times10^{-2}$ & 2.9$\times10^{-2}$  & 4.1$\times10^{-2}$\\[-0.15cm]
$^{110}$Xe$\rightarrow$$^{106}$Te+$\alpha$   &\hspace{0.4cm} 3.885        & 0.099 &  0.052 & 64\%       & 1.1$\times10^{-1}$ & 1.5$\times10^{-1}$  & 1.0$\times10^{-1}$\\[-0.15cm]
$^{111}$Xe$\rightarrow$$^{107}$Te+$\alpha$   &\hspace{0.4cm} 3.693        & 0.134 &  0.064 & $<$100\%   &$>$9.0$\times10^{-1}$& 2.6$\times10^{0}$  & 2.5$\times10^{0}$\\[-0.15cm]
$^{112}$Xe$\rightarrow$$^{108}$Te+$\alpha$   &\hspace{0.4cm} 3.330        & 0.134 &  0.056 & 0.8\%      & 3.4$\times10^{2}$  & 3.7$\times10^{2}$   & 3.8$\times10^{2}$\\[-0.15cm]
$^{113}$Xe$\rightarrow$$^{109}$Te+$\alpha$   &\hspace{0.4cm} 3.094        & 0.142 &  0.048 & 0.035\%    & 7.8$\times10^{3}$  & 3.3$\times10^{4}$   & 2.4$\times10^{4}$\\[-0.15cm]
$^{114}$Xe$\rightarrow$$^{110}$Te+$\alpha$   &\hspace{0.4cm} 2.775$^{*}$  & 0.152 &  0.049 &            &                    & 1.1$\times10^{7}$   & 2.2$\times10^{7}$\\[-0.15cm]
$^{115}$Xe$\rightarrow$$^{111}$Te+$\alpha$   &\hspace{0.4cm} 2.495$^{*}$  & 0.161 &  0.043 &            &                    & 1.1$\times10^{10}$  & 0.6$\times10^{10}$\\[-0.10cm]
\hline
$^{110}$Cs$\rightarrow$$^{106}$I+$\alpha$    &\hspace{0.4cm} 4.270$^{*}$  & 0.134 &  0.072 &            &                    & 1.5$\times10^{-2}$  & 0.8$\times10^{-2}$\\[-0.10cm]
$^{111}$Cs$\rightarrow$$^{107}$I+$\alpha$    &\hspace{0.4cm} 3.900$^{*}$  & 0.142 &  0.066 &            &                    & 6.8$\times10^{-1}$  & 9.6$\times10^{-1}$\\[-0.15cm]
$^{112}$Cs$\rightarrow$$^{108}$I+$\alpha$    &\hspace{0.4cm} 3.710$^{*}$  & 0.152 &  0.058 &            &                    & 1.6$\times10^{1}$   & 1.0$\times10^{1}$\\[-0.15cm]
$^{113}$Cs$\rightarrow$$^{109}$I+$\alpha$    &\hspace{0.4cm} 3.430$^{*}$  & 0.160 &  0.060 &            &                    & 5.1$\times10^{2}$   & 4.9$\times10^{2}$\\[-0.15cm]
$^{114}$Cs$\rightarrow$$^{110}$I+$\alpha$    &\hspace{0.4cm} 3.357        & 0.161 &  0.059 & 0.018\%    & 3.2$\times10^{3}$  & 2.9$\times10^{3}$   & 2.7$\times10^{3}$\\[-0.10cm]
\hline
$^{112}$Ba$\rightarrow$$^{108}$Xe+$\alpha$   &\hspace{0.4cm} 4.260$^{*}$  & 0.152 &  0.067 &            &                    & 2.0$\times10^{-2}$  & 1.0$\times10^{-2}$\\[-0.10cm]
$^{113}$Ba$\rightarrow$$^{109}$Xe+$\alpha$   &\hspace{0.4cm} 4.020$^{*}$  & 0.160 &  0.060 &            &                    & 5.4$\times10^{-1}$  & 7.6$\times10^{-1}$\\[-0.15cm]
$^{114}$Ba$\rightarrow$$^{110}$Xe+$\alpha$   &\hspace{0.4cm} 3.540        & 0.169 &  0.052 & 0.9\%      & 4.8$\times10^{1}$  & 2.6$\times10^{2}$   & 1.8$\times10^{2}$\\[-0.10cm]
\hline \hline
\end{tabular}
\footnotetext{Superscript * denotes the extrapolated or calculated $\alpha$-decay energies
of the predicted $\alpha$-emitters.}
\end{table*}

\end{document}